\documentclass[aps,pra,twocolumn,superscriptaddress]{revtex4}

\usepackage{amsmath,amssymb}
\usepackage{multirow}
\usepackage{xcolor}
\usepackage{srcltx}
\usepackage{hyperref,graphicx}

\def\be{\begin{equation}}
\def\ee{\end{equation}}
\newcommand{\ket}[1]{|#1\rangle}

\newcommand{\fisicarm}{Dipartimento di Fisica, Sapienza Universit\`{a} di Roma, Piazzale Aldo Moro, 5, I-00185 Roma, Italy}
\newcommand{\ino}{Istituto Nazionale di Ottica, Consiglio Nazionale delle Ricerche (INO-CNR), Largo Enrico Fermi, 6, I-50125 Firenze, Italy}
\newcommand{\cfermi}{Museo Storico della Fisica e Centro Studi e Ricerche Enrico Fermi, Via Panisperna 89/A, Compendio del Viminale, I-00184 Roma, Italy}
\newcommand{\ifn}{Istituto di Fotonica e Nanotecnologie, Consiglio Nazionale delle Ricerche (IFN-CNR), Piazza Leonardo da Vinci, 32, I-20133 Milano, Italy}
\newcommand{\fisicami}{Dipartimento di Fisica, Politecnico di Milano, Piazza Leonardo da Vinci, 32, I-20133 Milano, Italy}

\begin{document}

\title{Integrated photonic quantum gates for polarization qubits}
\author{Andrea Crespi}
\affiliation{\ifn}
\affiliation{\fisicami}

\author{Roberta Ramponi}
\affiliation{\ifn}
\affiliation{\fisicami}

\author{Roberto Osellame}
\affiliation{\ifn}
\affiliation{\fisicami}

\author{Linda Sansoni}
\affiliation{\fisicarm}

\author{Irene Bongioanni}
\affiliation{\fisicarm}

\author{Fabio Sciarrino}
\affiliation{\fisicarm}
\affiliation{\ino}

\author{Giuseppe Vallone}
\affiliation{\cfermi}
\affiliation{\fisicarm}

\author{Paolo Mataloni}
\affiliation{\fisicarm}
\affiliation{\ino}
%

\begin{abstract}
\end{abstract}

\maketitle

\textbf{
Integrated photonic circuits have a strong potential to perform quantum information processing \cite{ladd10nat,obri09np}. 
Indeed, the ability to manipulate quantum states of light by integrated devices may open new perspectives 
both for fundamental tests of quantum mechanics and for novel technological applications \cite{kok07rmp}. 
However, the technology for handling polarization encoded qubits, the most commonly adopted approach, 
is still missing in quantum optical circuits \cite{poli09ieee}. {Here we demonstrate the first integrated photonic Controlled-NOT (CNOT)
 gate for polarization encoded qubits. This result has been enabled by the integration, based on femtosecond laser waveguide writing, of partially polarizing beam splitters on a glass chip.} 
We characterize the logical truth table of the {quantum gate} demonstrating {its high fidelity to the expected one.}
{In addition, we} show the ability of this gate to transform separable states into entangled ones {and vice versa}. 
Finally, the full accessibility of our device is exploited to carry out a complete characterization 
of the CNOT gate through a quantum process tomography.
}

Incorporating the laws of quantum mechanics in data storage, 
processing and transmission may unleash new possibilities for information processing,
{ranging from quantum communication to quantum sensing, metrology, simulation and computing
\cite{ladd10nat,obri09np,kok07rmp,ursi07nap}}. 
In the last few years photonic quantum technologies have been adopted as a promising 
experimental platform for quantum information science\cite{kok07rmp}. 
The realization of complex optical schemes consisting of many elements requires the introduction of 
waveguide technology to achiethe desired scalability, 
stability and miniaturization of the device.  Recently, silica waveguide circuits on silicon 
chips have been employed in quantum applications to realize 
stable interferometers for two-qubit entangling gates\cite{poli08sci}. 
Such approach with qubits encoded into two photon optical paths, 
representing the logical basis $\{\ket{0},\ket1\}$, yielded the first demonstration of 
an integrated linear optical {CNOT} gate and enhanced quantum sensitivity in phase-controlled interferometers\cite{poli08sci,poli09sci, smit09ope, lain10apl}. 
However, many quantum information processes 
and sources of entangled photon states are based on the polarization 
degree of freedom \cite{walt05nat,ursi07nap}, which allows to implement quantum operations without the need of path duplication and thus with the simplest and most compact circuit layout.
Up to now, integrated devices able to efficiently guide and manipulate {polarization-encoded photonic qubits} are still lacking.

Femtosecond laser waveguide writing has been emerging as a powerful tool in the rapidly advancing field of quantum photonics\cite{mars09ope}. This technique, which has been developed {in recent} years, exploits nonlinear absorption of focused femtosecond pulses to induce permanent and localized refractive index increase {in} transparent materials\cite{gattass08np, dellavalle09joa, osellame03josab}. Waveguides are directly fabricated in the bulk of the substrate by translation of the sample at constant velocity with respect to the laser beam, along the desired path. This maskless and single-step technique allows fast and cost-effective prototyping of new devices, potentially exploiting also three-dimensional geometries impossible to obtain with conventional lithography. This technique is capable of producing high-quality waveguides with low birefringence, which are indeed suited for the propagation of polarization-encoded qubits, as we recently reported in \cite{sans10prl}, where a polarization insensitive directional coupler has been demonstrated.

In this Letter, we report on {the first integrated photonic CNOT gate for polarization encoded qubits
and characterize completely its quantum behavior. The demonstration of this quantum gate has been made possible by the} fabrication in a glass chip of integrated devices acting as partially polarizing beam splitters (PPBS). Precisely, we show that femtosecond laser writing enables direct {inscription} of directional couplers with fine and independent control on the splitting ratio for the horizontal (H) and vertical (V) {polarizations}. 

\begin{figure*}
\includegraphics[width=17cm]{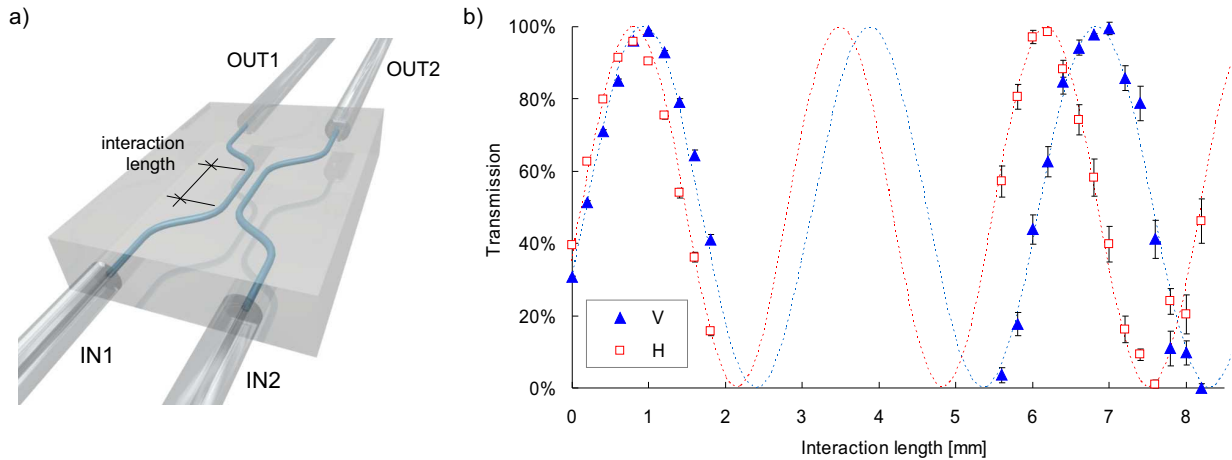}
\caption{\textbf{Partially polarizing directional couplers.} a) Schematic of a waveguide directional coupler. b) H (squares) and V (triangles) polarizations transmission of directional couplers with different interaction lengths, based on slightly birefringent waveguides. First, the 0-2 mm interaction length range was investigated to evaluate the beating length difference between the two polarizations; the interval of interest to obtain the required PPDCs was estimated to be in the 5.6-8.2 mm range, which was consequently explored. Transmission for 0 mm interaction length is non-zero, because some coupling already occurs in the curved portions of the approaching/departing waveguides. Error bars indicate fabrication reproducibility.}
\label{fig:chip}
\end{figure*}

Directional couplers are the integrated optical analogue of bulk beam splitters (BSs) and are thus fundamental building blocks of quantum optical circuits. In these devices (see schematic in 
Fig. \ref{fig:chip}a) two distinct waveguides are brought close together 
for a certain propagation length, called interaction length, so that the two propagating modes become coupled through evanescent field overlap. In analogy to bulk BSs, reflectivity and transmissivity ratios of the directional coupler can be conveniently defined: when light is launched into port IN1, referring to Fig. \ref{fig:chip}a, $R = P_{\mathrm{OUT 1}} / \left( P_{\mathrm{OUT 1}} +  P_{\mathrm{OUT 2}}\right)$ and $T = 1 - R = P_{\mathrm{OUT 2}} / \left( P_{\mathrm{OUT 1}} +  P_{\mathrm{OUT 2}}\right)$, respectively ($P$ indicates the optical power); the symmetry of the device guarantees that the same relations hold when light is launched into port IN2, by simply inverting the two indexes. Optical power transfer from one waveguide to the other follows a sinusoidal law with the interaction length, whose oscillation period (beating period) depends upon the coupling coefficient of the two guided modes according to coupled mode theory \cite{yariv73jqe}. If some waveguide birefringence is present, the coupling coefficient, and hence the beating period, can be different for the two polarizations (Fig. \ref{fig:chip}b). With high birefringence waveguides, this aspect has been exploited to implement polarizing beam splitters for telecom applications\cite{kiyat05ptl}. 

{However, high birefringence waveguides are not suitable for the propagation of polarization encoded qubits since they would cause decoherence between the photons typically generated by parametric down conversion, which are characterized by a large bandwidth $\Delta \lambda >$ 1 nm. The optimal waveguides for a polarization based quantum gate would therefore need to find the best compromise on the birefringence value, which should be low enough to preserve the coherence between photons, and high enough to enable polarization dependent processing. Such compromise is found in femtosecond laser written waveguides in a borosilicate glass. In fact, thanks to their low birefringence, these waveguides allow to propagate polarization encoded qubits without any perturbation and also to achieve practically polarization-insensitive devices, as demonstrated in \cite{sans10prl}. However, in this work we demonstrate that the same waveguides are also capable of performing polarization based processing of the qubits. In fact, if sufficiently long interaction lengths, covering a few beating periods, are implemented in directional couplers, it is possible to strongly enhance the difference in the polarization behavior of the device. This approach, as shown in Fig. \ref{fig:chip}b, allows to finely tailor the splitting ratios for the two polarizations and enables us to realize integrated PPBSs for the first time. It is important to note that the polarization sensitivity/insensitivity of the directional coupler can be selected by acting on geometrical parameters and not on physical properties of the waveguides; this allows simple and flexible design of complex integrated quantum devices.}

\begin{figure*}
\includegraphics[width=14cm]{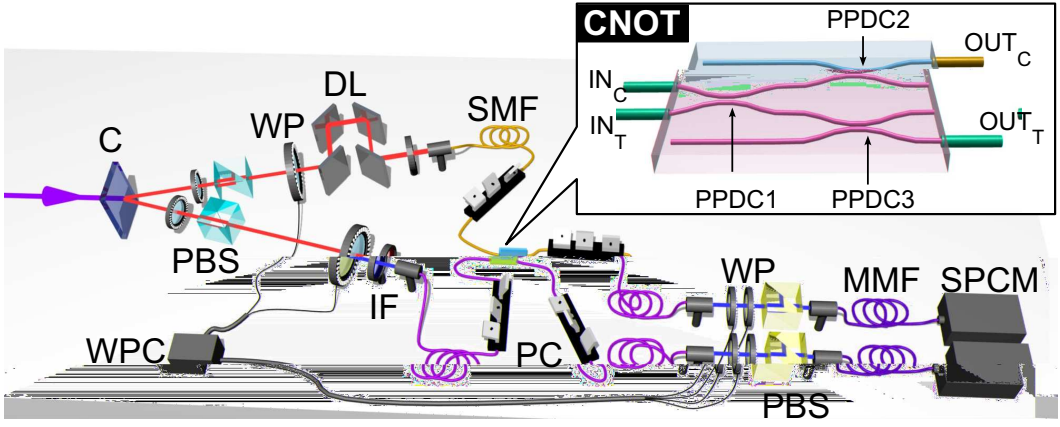}
\caption{
\textbf{Experimental setup.} Sketch of the experimental setup that can be conceptually divided into three parts.
(i) The source: 
{photon pairs at wavelength $\lambda=808nm$ were generated via
spontaneous parametric down conversion in a 1.5mm $\beta$-barium borate  
crystal (C) cut for type-II non-collinear phase matching, 
pumped by a CW laser diode with power $P = 50mW$ \cite{kwia95prl}.}
Photon polarization states are prepared by using polarizing beam splitters (PBSs) and waveplates (WPs). 
A delay line (DL) is inserted to control the temporal superposition of the photons, which are then coupled to single mode fibers (SMFs) and injected into the integrated CNOT gate. 
{Interference filters (IF) determine the photon bandwidth $\Delta\lambda=6nm$.}
(ii) Integrated CNOT for polarization encoded qubits (see inset) realized by ultrafast laser writing technique.
(iii) The analysis apparatus: the polarization state of qubits emerging from the chip is 
analyzed by standard analysis setups (WP+PBS). Photons are then delivered to single photon counting 
modules (SPCMs) through multimode fibers (MMFs) and coincidences between the two channels are measured.
Polarization controllers (PC) are used before and after the CNOT device to 
compensate polarization rotations induced by the fibers. A waveplate controller (WPC) drives 
the motorized waveplates to automatize the measurements.
}
\label{fig:setup}
\end{figure*}

{We will now demonstrate how {this technology} 
can be exploited to realize quantum optical gates.}
In the polarization-encoding approach, a generic qubit $\alpha\ket0+\beta\ket1$ is implemented by a coherent superposition
of H and V polarization states, $\alpha\ket H+\beta\ket V$, of single photons.
To achieve universal quantum computation, single qubit transformations together with a two-qubit gate are sufficient \cite{niel00QCQI}.
One-qubit logic gates are straightforwardly implemented by using birefringent waveplates.
The most commonly exploited two-qubit gate is the {CNOT that flips the target qubit (\textsf{T})
depending on the state of the control qubit (\textsf{C})}. The CNOT action is described by a unitary transformation acting on a generic superposition 
of two qubit quantum states. In the computational basis $\left\{\ket{00},\ket{01},\ket{10},\ket{11}\right\}$ for the systems \textsf{C} and \textsf{T},
the matrix associated to the CNOT is:
\begin{equation}
 \mathcal{U}_{CNOT}=
  \begin{pmatrix}
     1 & 0 & 0 & 0\\
     0 & 1 & 0 & 0\\
     0 & 0 & 0 & 1\\
     0 & 0 & 1 & 0\\
  \end{pmatrix}.
\label{Ucnot}
\end{equation}
A striking feature of this gate is given by the ability to entangle and disentangle qubits.
Precisely, the input states  ${\ket\pm}_\textsf{C}{\ket0}_\textsf{T}$ and ${\ket\pm}_\textsf{C}{\ket1}_\textsf{T}$, where $\ket{\pm}=\frac{1}{\sqrt2}(\ket0\pm\ket1)$, evolve into the following output states  $\ket{\Phi^\pm}=\frac{1}{\sqrt2}({\ket{0}}_\textsf{C}{\ket{0}}_\textsf{T}\pm{\ket{1}}_\textsf{C}{\ket{1}}_\textsf{T})$
and $\ket{\Psi^\pm}=\frac{1}{\sqrt2}({\ket{0}}_\textsf{C}{\ket{1}}_\textsf{T}\pm{\ket{1}}_\textsf{C}{\ket{0}}_\textsf{T})$, respectively, which are the so-called Bell states.
Photonic two-qubit gates need an interaction
between the two photons that carry the information, and this would suggest that a strong optical nonlinearity is required. However, it was demonstrated that scalable quantum computing is possible by using only 
linear optical circuits, mainly consisting of beam splitters \cite{knil01nat}. 
The most simple scheme to implement the polarization CNOT 
exploits three PPBSs with suitable polarization dependent transmissivites \cite{ralp02pra,hofm02pra}.
The two-photon interaction occurs by a Hong-Ou-Mandel effect\cite{hong87prl} in a single PPBS, 
while the other two PPBSs operate as compensators. 
Up to now such scheme has been experimentally implemented only using bulk optics \cite{kies05prl2,okam05prl,lang05prl}.
The integrated CNOT gate is achieved by 
the optical scheme given in the inset of Fig. 2:
it consists of a  first partially polarizing directional coupler (PPDC1), with
transmissivities $T_H^{(1)}=0$ and $T_V^{(1)}=\frac{2}{3}$, 
where target and control qubits interfere, followed by two other 
directional couplers (PPDC2 and PPDC3), with ${T_H^{(2,3)}}= \frac{1}{3}$ and ${T_V^{(2,3)}}= 1$,
where the horizontal and vertical polarization contributions are balanced.
In this scheme, the following correspondence between logical qubits and physical states holds: 
$\ket{0}_\textsf{C}\equiv\ket V_\textsf{C}$, $\ket{1}_\textsf{C}\equiv\ket H_\textsf{C}$,
$\ket{0}_\textsf{T}\equiv\ket{A}_\textsf{T}$, $\ket{1}_\textsf{T}\equiv\ket{D}_\textsf{T}$ where
$\ket{A}=\frac{1}{\sqrt2}(\ket{H}+\ket{V})$ and $\ket{D}=\frac{1}{\sqrt2}(\ket{H}-\ket{V})$. The CNOT operation succeeds with probability $P=\frac{1}{9}$.

The PPDCs have been fabricated by the femtosecond laser waveguide writing technique above described.
In order to calibrate the fabrication parameters, several directional couplers have been produced with different interaction lengths ranging from $0 mm$ to $2 mm$ and from $5.6 mm$ to $8.2 mm$ (see Fig. 1). The distance between the two waveguides in the interaction region was kept constant at $7 \mu m$. It can be observed that for an interaction length $L_1 \simeq 7.4 mm$ a device with $T_H = 0$ and $T_V = \frac{2}{3}$ is obtained, fulfilling the requirements for PPDC1, whereas a length $L_2 \simeq 7 mm$ provides $T_H= \frac{1}{3}$ and $T_V=1$ and can be adopted for PPDC2 and PPDC3.
Several CNOTs have been fabricated according to the schematic of Fig. 2 (inset) with slightly different interaction lengths around the values of $L_1$ and $L_2$ to take into account possible fabrication imperfections. The footprint of each integrated CNOT is $500 \mu m \times 3 cm$.
After characterization with a classical laser source, the devices providing the best estimated performance have been selected. The device used in the experiments has the following parameters: {$\overline{T}_H^{(1)} < 1 \% $, $\overline{T}^{(1)}_V= (64\pm1) \% $, $\overline{T}_H^{(2)}= (43\pm1) \% $, $\overline{T}_V^{(2)}= (98\pm1)\% $,$\overline{T}_H^{(3)}= (27\pm1)\% $, and $\overline{T}_V^{(3)}= (93\pm1)\% $. }
To improve the robustness and stability of the device, single-mode fiber arrays have been permanently bonded to the input and output ports.

\begin{figure}
\includegraphics[width=8cm]{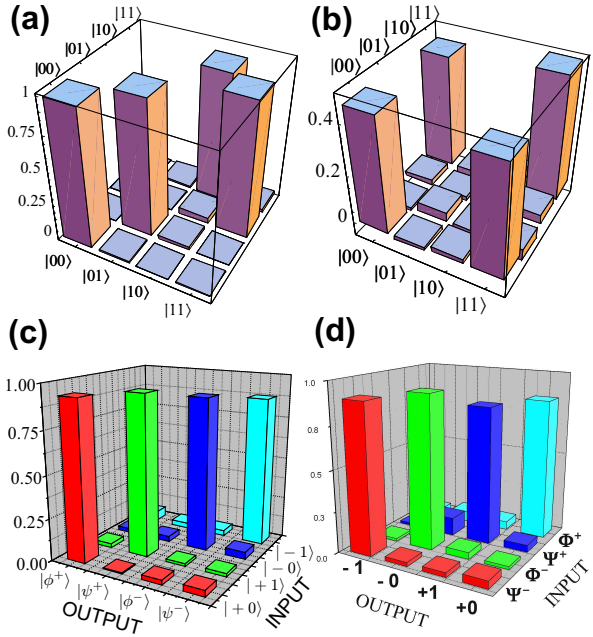}
\caption{\textbf{Truth table and entanglement generation.} \textbf{(a)} Truth table obtained after correction of photon distinguishability. \textbf{(b)} Experimental tomographic reconstruction of 
 the Bell state $\ket{\Phi^+}$ obtained by injecting the separable state $\ket{+0}$ in the CNOT. \textbf{(c)} Measured probabilities of the output Bell states corresponding to the different input separable states.
 \textbf{(d)} Measured probabilities of the output separable states corresponding to the different input Bell states.}
\label{fig:truetable}
\end{figure}

As a first quantum experiment we determined the truth table of the device operated as a CNOT gate. 
On this purpose we used the experimental setup represented in Fig. \ref{fig:setup}. Temporal superposition of photon wavepackets in PPDC1 was obtained by acting on the delay line DL. Then 
we injected into the chip the four computational basis states $\ket{0}_\textsf{C}\ket{0}_\textsf{T}$,
$\ket{0}_\textsf{C}\ket{1}_\textsf{T}$,
$\ket{1}_\textsf{C}\ket{0}_\textsf{T}$ and
$\ket{1}_\textsf{C}\ket{1}_\textsf{T}$ 
and measured the probability of detecting each of them at the output. 
The average measured fidelity of the logical basis \cite{kies05prl2} has been calculated as $\mathcal{F}=0.940\pm0.004$. A partial distinguishability of photon wavepackets, measured by two-photon Hong-Ou-Mandel interference \cite{hong87prl}, 
slightly reduces the fidelity of the gate.
By correcting the truth table for such imperfection, we obtain the result reported in Fig. \ref{fig:truetable}-(a), 
which provides the actual fidelity of the integrated device $\mathcal{F_{\mathrm{dev}}}=0.970\pm0.008$ (see Methods). We can compare this value with the expected fidelity of the device {$\overline{\mathcal{F}}=0.975\pm0.007$}, estimated by taking into account the measured transmissivities of the PPDCs. 

As already mentioned, the CNOT can also be exploited as an entangling gate. 
We experimentally verified this behavior by injecting into the device 
the states $\left\{\ket{\pm}_\textsf{C}\ket{0}_\textsf{T},\ket{\pm}_\textsf{C}\ket{1}_\textsf{T}\right\}$ and measuring a set of 
observables in order to obtain a tomographic reconstruction of the density matrices of the output states. 
As expected the action of the CNOT converts the separable states into the maximally entangled Bell states.
We report in Fig. \ref{fig:truetable}-(b) one of the reconstructed density matrices and in Fig. \ref{fig:truetable}-(c) the  probabilities to generate the different Bell states.
The fidelities of the output density matrices compared with those of the corresponding Bell states are {$\mathcal{F}_{\ket{\Phi^+}}=0.930\pm0.014,\mathcal{F}_{\ket{\Psi^+}}=0.939\pm0.
008,\mathcal{F}_{\ket{\Phi^-}}=0.900\pm0.006,\mathcal{F}_{\ket{\Psi^-}}=0.877\pm0.011$, with average value ${\mathcal{F}}_{Bell}=0.912\pm0.005$.
Note that, no photon distinguishability correction was performed to obtain Bell state density matrices.
We also checked that the CNOT gate can be adopted to discriminate the four Bell states.
Indeed, the four entangled states $\ket{\Psi^\pm}$, $\ket{\Phi^\pm}$ are transformed into four orthogonal separable
states which can be easily discriminated, as shown in Fig. \ref{fig:truetable}-(d). The discrimination probability
is $0.877\pm0.007$, slightly lower than the previous fidelities due to imperfections in the entanglement source.

\begin{figure}
\centering
\includegraphics[width=8.5cm]{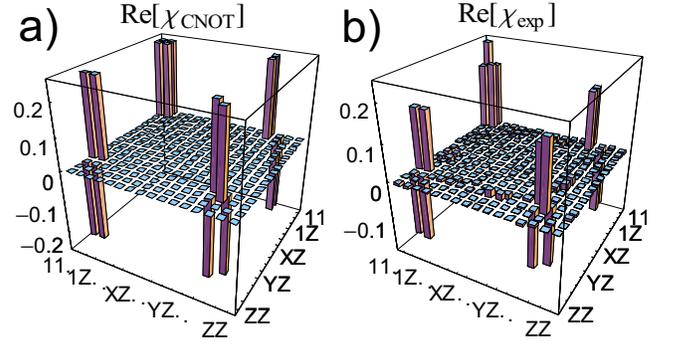}
\caption{a) Ideal $\chi_{CNOT}$ and b) experimental $\chi_{exp}$ matrices (real part) obtained from the quantum process tomography of the CNOT gate.
Imaginary part of $\chi_{exp}$ is negligible. $X$, $Y$ and $Z$ correspond to Pauli matrices
$\sigma_1$, $\sigma_2$ and $\sigma_3$, respectively.}
\label{chi}
\end{figure}
{To provide a full characterization of the quantum device we carried out 
a quantum process tomography \cite{chua97jmo}.
This involves preparing the photons in a complete set of input basis states, 
and characterizing the output.
A generic quantum process $\mathcal{E}$ acting on 2-qubit density matrix $\rho$,
can be expressed as $\mathcal{E}(\rho)=\sum^{15}_{m,n=0}\chi_{mn}\Gamma_m\rho \Gamma_n^{\dagger}$
where the operators $\Gamma_m$ are defined as tensor product of Pauli matrices 
$\left\{\Gamma_m\equiv\sigma_i\otimes\sigma_j\right\}$, $i,j=0,\dots,3$, $m=0,\dots,15$.
The matrix $\chi_{mn}$ contains all the information of the process. 
We reconstructed the 16$\times$16 matrix $\chi_{exp}$ shown in Fig. \ref{chi}).
The process fidelity \cite{bong10pra} (see Methods)
with the ideal CNOT gate was measured as $\mathcal{F}_{exp}=0.906\pm0.003$.
As already done for the truth table, we corrected the contribution due to distinguishable photons
obtaining a fidelity ${\mathcal{F}}=0.943\pm0.006$.}

In summary we reported on {the integration of partially polarizing beam splitters on a glass chip, enabling the demonstration of the first integrated {photonic CNOT gate based on polarization encoding. From the fabrication point of view this work shows the capability of femtosecond laser microfabrication to produce also polarization sensitive waveguide devices, thus further enriching the portfolio of applications that can be addressed by this simple and flexible fabrication technique.}}
This work represents a major step towards the development of integrated photonic technology which could provide a viable solution for quantum information processing {and paves the way to the integration of a wealth of polarization based quantum algorithms developed for bulk optical circuits.}
Future researches will be devoted to the {on chip} implementation { of both pre-oriented and tunable waveplates in order
to also integrate one qubit gates}. 
{The present results
open new perspectives towards joint integrated handling of hybrid quantum states}
based on different degrees of freedom of light \cite{ross09prl, chiu10prl, gao10prl, naga09npho,naga10prl,barr08nap},
 such as polarization, path and orbital angular momentum.

\section*{Aknowledgements}
{This work was supported by FIRB-Futuro in
Ricerca HYTEQ, FARI project 2010 of Sapienza Universit\`a di Roma,
and Finanziamento Ateneo 2009 of
Politecnico di Milano.

\section*{Correspondence} Correspondence and requests for materials should be addressed to
Fabio Sciarrino (email: fabio.sciarrino@uniroma1.it) and Roberto Osellame ({roberto.osellame@ifn.cnr.it}).

\section*{Methods}

\textbf{Chip fabrication and characterization} - Waveguides were micromachined in borosilicate glass substrate (Corning EAGLE2000) using a commercial HighQ FemtoREGEN femtosecond laser, which provides $400 fs$ pulses at $960 kHz$ repetition rate. For the waveguide fabrication pulses with $240 nJ$ energy were focused $170 \mu m$ under the glass surface, using a $0.6 N.A.$ microscope objective while the sample was translated at a constant speed of $20 mm/s$ by high precision, three-axis air-bearing stages (Aerotech FiberGlide 3D).
The guided mode at $806 nm$ is slightly elliptical, measuring $8 \mu m \times 9\mu m$. Measured propagation losses are $0.8 dB/cm$ and coupling losses to single mode fibers about $1.3 dB$ per facet. Birefringence of these waveguides is on the order of $B = 7 \times 10^{-5}$, as characterized in \cite{sans10prl}.
For the curved segments in the directional couplers of the integrated CNOTs a $30 mm$ bending radius was adopted, which gives less than $1 dB$ of additional bending losses on the whole devices.

\textbf{Photon distinguishability} -
In order to evaluate the imperfections caused by a partial photon
distinguishability, we performed a Hong-Ou-Mandel
experiment by using the same parametric source and a single mode  fiber beam splitter. 
By the measured interference visibility it is possible to 
infer the photon distinguishability, mainly caused by a spectral bandwidth mismatch. We can define a distinguishability parameter $p$ which
is directly related to the ratio between the measured and theoretical visibility, 
{(depending on the fiber beam splitter reflectivity)}
$1-p=\frac{V_{meas}}{V_{theo}}$. For our setup we can estimate $p=0.030\pm0.005$.
This parameter has been used to correct the CNOT truth table. Precisely, the correction has been performed by measuring how the input states evolve through the CNOT device when the two photons are distinguishable, a condition achieved by a temporal mismatch of the photons at PPDC1.
{Then, the contribution to the output state probabilities due to photon
distinguishability, has been subtracted with weight $p$
from the measured truth table of figure 3-(a).}

{\textbf{Quantum process tomography -}
The full characterization of a quantum device is provided by quantum process tomography \cite{chua97jmo}.
Following the procedure adopted in Ref. \cite{bong10pra} we performed the quantum process tomography of our device and reconstructed the associated $\chi$ matrix. The experimental setup is the same adopted for the previous measurements (see Fig. \ref{fig:setup}): we prepared 16 different input states
and measured the projections of the output states on the set $\left\{\ket{ij}\right\}$, where $i,j=H,V,D,A,R,L$, i.e. horizontal, vertical, diagonal,
antidiagonal, righthanded and lefthanded polarization, using standard polarization analysis setups. The obtained $\chi$ matrix is reported in
Fig. \ref{chi}. In order to evaluate the quality of the acquired data we calculated the process fidelity \cite{bong10pra} as
\begin{equation}
 \mathcal{F}_{exp}=
 \frac{\mathrm{Tr}\left[\sqrt{\sqrt{\chi_{exp}}\chi_{CNOT}\sqrt{\chi_{exp}}}\right]^2}{\mathrm{Tr}[\chi_{exp}]\mathrm{Tr}[\chi_{CNOT}]},
\label{fidchi}
\end{equation}
where $\chi_{CNOT}$ is the process matrix of the ideal CNOT gate. 
}

%

\end{document}